# Nonlinear Characteristics of Neural Signals

Z.Zh. Zhanabaev[a,b], T.Yu. Grevtseva[a,b], Y.T. Kozhagulov[c,*]

[a] National Nanotechnology Open Laboratory, Al-Farabi Kazakh National University, al-Farabi Avenue, 71, Almaty 050038, Kazakhstan
Tatyana.Grevtseva@kaznu.kz

[b] Scientific-Research Institute of Experimental and theoretical physics, Al-Farabi Kazakh National University, al-Farabi Avenue, 71, Almaty 050038, Kazakhstan
kaznu.kz@list.ru

[c] Department of Solid State Physics and Nonlinear Physics, Faculty of Physics and Technology
al-Farabi Kazakh National University, al-Farabi Avenue, 71
Almaty 050038, Kazakhstan
kazgu.kz@gmail.com

ABSTRACT

The study is devoted to definition of generalized metrical and topological (informational entropy) characteristics of neural signals via their well-known theoretical models. We have shown that time dependence of action potential of neurons is scale invariant. Information and entropy of neural signals have constant values in case of self-similarity and self-affinity.

## 1. Introduction

Study of artificial neural networks is one of the most urgent problems in information technology. Generally, neural networks should have the basic properties of an ensemble of biological neurons: associativity, resistance to noise, distributed nature of information storage and processing, adaptability to relationships between elements. This was reported in many reviews and articles [1-4]. Results of recent researches indicate to the possibility for using of neural oscillations for routing of information [5-7].

However, using of biological neuron model (for example, ion channel functioning model of a neuron) involves analysis of great number of systems of differential equations containing big number of parameters with the same order [8-10]. Thus, it's necessary to find ways for establishment of universal, the most common and simple laws of neuron dynamics. Application of methods of nonlinear dynamics for the description of neuron networks by use of systems of differential equations are presented in [11,12].

There are certain theoretical models [13-16] describing the following regularities. A neuron moves from the stable state to an excited state because of external stimulus (external field). This process is characterized by appearance of spikes and bursts. A neuron also can generate quasi-periodic, chaotic and noise-like oscillations [16], along with the bursting oscillations.

Nonlinear dynamics [17] describes chaotic phenomena on the base of fractal and informational-entropic laws. Informational entropy and fractal dimensions of physical quantities set are used as quantitative characteristics of chaos [18,19]. A possibility to describe different types of behaviors of neuronal action potentials via fractal measures is described in [16]. So, we can formulate a problem: do any entropic laws describing neurons dynamics exist? Obviously, in case of self-similarity (similarity coefficients for the different variables are equal each other) and self-affinity (similarity coefficients are different) of fractal measures we expect to find the existence of intervals with constant entropy. Formulation of two additional laws (fractal and entropic) would help us to build a more efficient neural networks for classification, recognition, identification, and processes control.

Purpose of the present work is to establish informational-entropic laws for the description of neural oscillations.

## 2. Equations for action potentials of neurons

There are various models of neural oscillations [13-16].

The FitzHugh-Nagumo model [13] consists of two equations, one governing a voltage-like variable $v$ having a cubic nonlinearity and a slower recovery variable $w$. It can be written as:

$$\dot{v} = v - v^3/3 - w + I_{ext}$$
$$\tau \cdot \dot{w} = v + a - b \cdot w.$$
(1)

The parameter $I_{ext}$ models the input current the neuron receives; the parameters $a$, $b > 0$ and $\tau > 0$ describe the kinetics of the recovery variable $w$.

The Hindmarsh-Rose models [14] equation can be written as:

$$\dot{x} = y - a \cdot x^3 + b \cdot x^2 - z + I$$
$$\dot{y} = c - d \cdot x^2 - y \qquad (2)$$
$$\dot{z} = r \cdot [s \cdot (x - x_R) - z],$$

where: $I$ mimics the membrane input current for biological neurons; $b$ allows one to switch between bursting and spiking behaviors and to control the spiking frequency; $\mu$ controls the speed of variation of the slow variable $z$, $s$ governs adaptation: a unitary value of $s$ determines spiking behavior without accommodation and subthreshold adaptation, $x_R$ sets the resting potential of the system.

Modeling of spiking-bursting neural behavior is possible via two-dimensional map described in [15]:

$$x_{i+1} = f(x_i, y_i)$$
$$y_{i+1} = y_i - \mu \cdot (x_i + 1) + \mu \cdot \sigma$$
$$f(x, y) \begin{cases} \alpha/(1-x) + y; & x \leq 0 \\ \alpha + y; & 0 < x < \alpha + y \\ -1; & x \geq \alpha + y, \end{cases} \qquad (3)$$

where $x_n$ is the fast and $y_n$ is the slow dynamical variable. Slow time evolution of $y_n$ is due to small values of the parameter $\mu = 0.001$. The term $\sigma_n$ can be used as the control parameter to select the regime of individual behavior. $\alpha$ is a control parameter of the map.

In the work [16] action potentials of neurons were taken in the form of fractal measures. We consider a fractal measure as a physical quantity characterized by an additive and a measurable set. As opposite to the well-known theories of fractals we choose value of scale of measurement not randomly, but as a relative difference between the desired and the external (control) parameter. Hence, the fractal measure is a nonlinear function depending on the process (object).

The traditional definition of fractal measure $M$ can be written as

$$M = M_0 \left( |\Delta M| / M_* \right)^{-\gamma}, \gamma = D - d, \gamma > 0, \qquad (4)$$

where $M_0$ is a regular (non-fractal) measure, $\Delta M$ is a scale of measurements, $M_*$ is norm of $M$, $D$ is fractal dimension of the set of values of $M$, $d$ is topological dimension of norm carrier. $\Delta M$ is independent on $M$, therefore, measure defining by (1) can be tentatively called the linear value. The dependence of M from A assumes the existence of certain conditions in the form of external disturbance, generally it is a order parameter.

Let us consider $\lambda$ as parameter of order. So, we can choose $\Delta M$ as

$$\Delta M_M = \frac{|M - \lambda|}{|M|} = \left| 1 - \frac{\lambda}{M} \right|, \qquad (5)$$

$$\Delta M_\lambda = \frac{|M - \lambda|}{|\lambda|} = \left| 1 - \frac{M}{\lambda} \right|, \qquad (6)$$

where indexes $M$ and $\lambda$ correspond to the norms $\Delta M$, taken as $M_* = |M|, |\lambda|$. According to (5) and (6) we can rewrite the formula (4) as

$$M_M = M_0 \left( \left| 1 - \frac{\lambda}{M} \right| \right)^{-\gamma}, \qquad (7)$$

$$M_\lambda = M_0 \left( \left| 1 - \frac{M}{\lambda} \right| \right)^{-\gamma}. \qquad (8)$$

At $\gamma \to \infty$ we have $M_M = M_\lambda = M_0$, it corresponds to the meaning of $M_0$. At $\lambda = 0$ we have $M_M = M_0, M_\lambda = 0$. It means that the fractal measure defined by its own norm exists in a case when external influence characterized by parameter $\lambda$ is absent.

Let us apply the formula (7) and (8) for the description of the action potential of neurons. We use the simplified notations of $M_M = M = V$, $M_0 = V_0$, $\lambda = F(t)$, $M_\lambda = V_F$. From (7), (8) we obtain the expressions for the potential $V$ of a neuron as

$$V = f(V_0, V), f(V_0, V) = V_0 \left( \left| 1 - \frac{F(t)}{V} \right| \right)^{-\gamma}, \qquad (9)$$

$$V = f(V_{0,F}, V_F), f(V_{0,F}, V_F) = V_{0,F} \left( \left| 1 - \frac{V_F}{F(t)} \right| \right)^{-\gamma}, \qquad (10)$$

where $V_0, V_{0,F}$ are the threshold excitation potentials. Neurons have inherent properties of quasi-particles, they can't exist without movement and outside of medium, and can be considered as fluctuation of the medium. Neurons communicate by sending action potentials. Therefore it is natural to assume that the action potentials has modulation - periodic nature, and in equations (9), (10) we can take the external field in the form

$$F(t) = A(1 + B \sin(\Omega t)), \qquad (11)$$

where $A, B, \Omega$ is amplitude, coefficient (deeps of modulation), frequency modulation of neural oscillations.

For the system consisting of $N$ neurons, we can write the following equations in iterative form as

$$V_{i+1}^{(k)} = V_0^{(k)} \left( \left| 1 - F^{(k)}(t) \middle/ \sum_{k=1}^{N} V_i^{(k)} \right| \right)^{-\gamma_k}, \quad (12)$$

$$V_{i+1}^{(k)} = V_0^{(k)} \left( \left| 1 - \sum_{k=1}^{N} V_i^{(k)} \middle/ F^{(k)}(t) \right| \right)^{-\gamma_k}, \quad (13)$$

where $k$ is number of neurons, $V_i$ is action potential of neurons, $V_0$ is threshold excitation potentials, $F(t)$ – modulated stimulus value of one neuron, $\gamma_k$ is the difference between fractal and topological dimension of the set of values $V_i$. Equation (12) takes into account a possibility for own sub-threshold neuron oscillations at $F(t) = 0$, and (13) is used only for the description of presence of the stimulus $F(t) \neq 0$. $F(t)$ can be caused by action potentials of neighboring neurons.

Equations (12) and (13) describe experimentally observed variety of spikes, chaotic oscillations, phase synchronization after the bursts [16].

## 3. Informational – entropic characteristics of neural signals

Conception of information is widely used in cybernetics, genetics, sociology, and so on. Development of open system physics stimulates formulation of universal definition of information which can be used in different branches of science. Definition of open system contains conception of information: open system is a system in which energy, matter and information are exchanged with its environment.

As usual, definition of a complex object includes list of its main properties. Information $I(x)$ for statistical realization of a physical value $x$ is greater than zero and can be defined for a non-equilibrium state ($I(x) \neq I(x_0)$ if $x \neq x_0$). Let us consider that $P(x)$ is probability of realization of the variable $x$. So, quantity of information can be described as

$$I(x) = -\ln P(x). \quad (14)$$

Reiteration and non-equilibrium character of a process can be taken into account by the condition $0 < P(x) < 1$. A lot of definitions of information have been suggested in different branches of science, but (14) corresponds to all of them.

Information can be defined as

$$I(x/y) = S(x) - S(x/y), \quad (15)$$

where $S(x)$ is absolute information entropy of an event $x$ and $S(x/y)$ is conditional entropy of an event $x$ when another event $y$ is to have occurred. Equation (15) can be used for solving of technical problems such as for estimation of transmission capacity (in communication channels). Informational entropy which is Shannon entropy $S(x)$ can be defined as mean value of information as

$$S(x) = \sum_i P_i(x) I_i(x) = -\sum_i P_i(x) \ln P_i(x). \quad (16)$$

Here, $i$ is number of a cell after segmentation of $x$. So, let us use (14) as a main definition of information.

According to (16) value of entropy calculated via probability density tends to infinity if $x$ is a continuous value. Let us try to define scale-invariant regularities. So, we have to use a new approach for the description of information phenomena. Because of this fact we can use information as a independent defining variable. Statistical characteristics of a process can be described via information. So, we can try to find new properties of information independent on scale of measurement.

Therefore, according to (14) we shall describe probability of realization of information $P(I)$ as

$$P(I) = e^{-I}. \quad (17)$$

Probability function $f(I)$ can be defined via the following relations:

$$0 \leq P(I) \leq 1, 0 \leq I \leq \infty; \int_0^\infty f(I) dI = 1,$$
$$P(I) = \int_I^\infty f(I) dI, f(I) = P(I) = e^{-I}. \quad (18)$$

Probability function $P(I)$ equals to density of probability distribution function $f(I)$. Information defined via (14) is characterized by property of scale invariance. It means that a whole object and its parts have the same law of distribution. Informational entropy $S(I)$ of distribution of information can be defined as mean value of information as

$$S(I) = \int_I^\infty I f(I) dI = (1 + I) e^{-I}. \quad (19)$$

Values of entropy can be normalized to unit, so, for $0 \leq I \leq \infty$ we have $1 \geq S \geq 0$. It is well-known that entropy of a continuous set tends to infinity at jumping values of variables. Therefore, we must calculate the integral by use of the Lebesgue measure. Describing information as the measure we have got (19).

Let us consider that a scale-invariant function $g(x)$ justifies to the well-known functional equation as

$$g(x) = \alpha g(g(x/\alpha)), \quad (20)$$

where $\alpha$ is scaling factor. All continuous functions in theirs fixed points justify (20). We use $f(I)$ and $S(I)$ as characteristic functions. Fixed points of the functions are [20]:

$$f(I) = I, e^{-I} = I, I = I_1 = 0.567, \quad (21)$$

$$S(I) = I, (1+I)e^{-I} = I, I = I_2 = 0.806. \quad (22)$$

The fixed points are limits for the following infinite maps

$$I_{i+1} = f(I_i), \lim_{i \to \infty} \exp(-\exp(... - \exp(I_0)...)) = I_1 \quad (23)$$

$$I_{i+1} = S(I_i), \lim_{i \to \infty} \exp(-\exp(... - \exp(\ln(I_0 + 1) - I_0)...)) = I_2 \quad (24)$$

at any initial values $I_0$, shown in Fig. 1. Number of brackets equals to $i+1$.

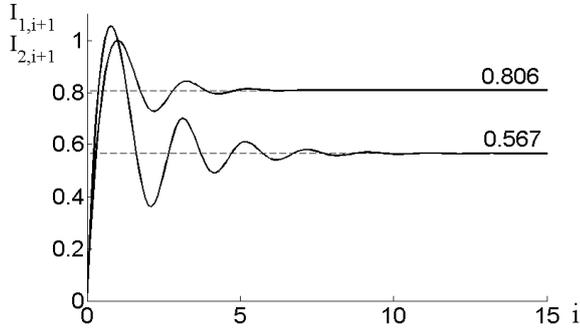

**Fig. 1.** Establishment of fixed values of information and entropy.

Interpretations of physical meaning of numbers $I_1 = 0.567$ and $I_2 = 0.806$ can be different. Probability density is a local (instant) characteristic. Therefore, it can be different for different variables. So, $I_1$ can be used as a criterion of self-affinity. Entropy is an averaged characteristic. Therefore, $I_2$ can be considered as a criterion of self-similarity.

On the other hand, numbers $I_1$ and $I_2$ can be considered as analog of the Fibonacci number $I_{20} = 0.618$ ("golden section") of dynamical measure) for statistical self-affine and self-similar systems correspondently. From (22) at $I \lesssim 1$ we have

$$(1+I)(1-I) = I, I^2 + I - 1 = 0, I = I_{20} = 0.618, \quad (25)$$

at $I \ll 1$ from the same equation we have $e^{-I} = I, I = I_1$. Therefore, we can use only (22) for the description of regularities of self-affinity, dynamical equilibrium and self-similarity of dynamical systems.

## 4. Generalized metrical characteristic of neural signals

In Part 3 of the present work we have described informational-entropic characteristics of fractal (non-Euclidean) signals. Let us consider generalization of metrical characteristic for the description of fractal (neural) signals. The generalized metrical characteristic is followed from the well-known Hölder's inequality for different functions [21, 22].

$$\left(\frac{1}{T}\int_0^T |x_i(t)|^p dt\right)^{1/p} \left(\frac{1}{T}\int_0^T |x_j(t)|^q dt\right)^{1/q} \leq$$
$$\leq K_{x_i,x_j}^{p,q} \frac{1}{T}\int_0^T |x_i(t) x_j(t)| dt, \frac{1}{p} + \frac{1}{q} = 1 \quad (26)$$

where $K_{x_i,x_j}^{p,q}$ is a coefficient, at constant value of this coefficient we have equilibrium in (26), $x_i(t), x_j(t)$ are physical values depending on time $t$, $T$ is characteristic time sufficient to establishment of statistical regularities, $p$ and $q$ are parameters of the system. At $p=q=2$ value of $K_{x_i,x_j}^{2,2}$ characterizes Euclidean metric of a set of values of functions $x_i(t), x_j(t)$. At $x_j(t) = 1$ we have form factor of the signal $x(t) \equiv x_i(t)$ given as

$$K_{x(t)} = \frac{\left(\langle x^2(t) \rangle\right)^{1/2}}{\langle |x(t)| \rangle}. \quad (27)$$

Here we have averaged values over the system. Parameter $K_{x(t)}$ is used in radio engineering. Corresponding choice of values of $p,q$ let us to use $K_{x_i,x_j}^{p,q}$ for the description of fractal signals. Let us designate $D$ as fractal dimension of the curve $x(t)$. So, we can express $p = D, q = D/(D-1)$. Using $x_i = x(t), x_j = t$ we have

$$K_{x,t}^{D,q} = \frac{\left(\langle |x|^D \rangle\right)^{1/D} \left(\langle |x_j|^q \rangle\right)^{1/q}}{\langle |x \cdot t| \rangle}, q = D/(D-1). \quad (28)$$

So, values of $K_{x,t}^{D,q}$ for signals with different coefficients of similarity $x,t$ can be used for distinguishing of various self-affine neural oscillations.

## 5. Metric-topological diagram of neural signals

Dependence of normalized informational entropy $S(x)$ (calculated via (16)) of neural signals belongs to different types of corresponding values of metrical characteristic $K_{x,t}^{D,q}$ (calculated via (28)) is shown in Fig. 2. Entropy has been defined via probabilities of entering of values (action potential of neurons) to the range $\delta = x_{i+1} - x_i, i = 1, 2, ...$, number of counts is $j = 1, ...N$, $N = 10^3$; $\delta = 10^{-4} \div 10^{-2}$. Let us consider impulses with different shape. Entropy of an impulse with shape of isosceles triangle is maximal because the distribution $x_i(t_j)$ is uniform (linear). So, we have choose $S_\Delta$ as a norm of entropy. $S(\delta)$ increases at decreasing of $\delta$, but value $S(\delta)/S_\Delta(\delta)$ in the selected range is practically constant.

$K \equiv K_{x,t}^{D,q}$ contains fractal dimension $D$ of curve $x(t)$. Value of the fractal dimension can be defined via the following expression [17] as

$$D = \lim_{\delta \to 0} \frac{\ln C(\delta)}{\ln 1/\delta},$$

$$C(\delta) = \lim_{N \to \infty} \frac{1}{N^2} \sum_{\substack{i=1 \\ i \neq j}}^{N} \sum_{j=1}^{N} \theta(\delta - |x_i - x_j|). \quad (29)$$

For clarity, each region of the diagram $S(K)$ contains examples of neural signals according to corresponding models ( □ – FitzHugh-Nagumo model [13], △ – Hindmarsh-Rose model [14], ○ – modeling of spiking-bursting neural behavior using two-dimensional map [15], ∗ – general model of scale invariant model of neural networks [16] ).

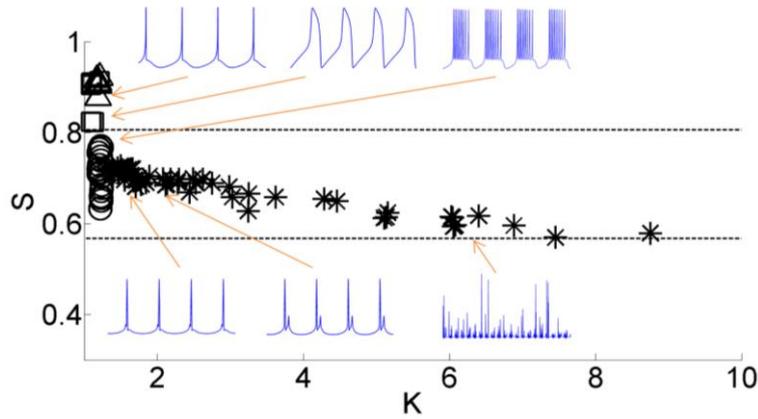

**Fig. 2.** Dependence of normalized entropy on generalized metrical characteristic for models of neural signals ( □ – FitzHugh-Nagumo model (1) $a = 1.5, b = 1, \tau = 12.5, I \in [1 \div 2]$, △ – Hindmarsh-Rose model (2) $a = 3, b = -1, c = 1, d = 5, x_R = 1.6, r = 0.005, s = 4, I \in [1.3 \div 3]$, ○ – modeling of spiking-bursting neural behavior using two-dimensional map (3) $\mu = 0.001, \sigma = 0.14, \alpha \in [4 \div 6]$, ∗ – general model of scale invariant model of neural networks (12) $A = 0.8, B = 0.8, V_0 = 0.1, \gamma \in [0.433 \div 1]$ ).

As follows from the diagram $S(K)$, the scale-invariant signals described via the fractal model are exactly self-affine signals (with big values of $K$), and normalized entropy has big values in the range $I_1 < S < I_2$ of fixed points $S$ and $I$. At $1 < K \leq 2$ oscillations are close to stochastic, and their entropy $S \geq I_2$.

In case of sufficient resolution $(\delta, N)$ the dependence $S(K)$ can be used for unambiguous identification of neural signals and for a criterion of solution for output layer of a neural network.

## 6. Conclusions

In this work we have shown the possibility for classifying of neural signals according to their nonlinear (entropic and generalized-metric) characteristics. We have defined stable, fixed points of neural signals which can be considered as criteria of self-affinity and self-similarity. Theoretical analysis of experimental data shows that neural signals are characterized by self-affine regularities.

**Acknowledgments**
The author declares that there is no conflict of interest regarding the publication of this article.